\documentclass[prd,amsmath,amssymb,floatfix]{revtex4}

\usepackage{graphicx}

\begin{document}

\title{Toward homochiral protocells in noncatalytic peptide systems}

\author{Marcelo Gleiser}
\email{gleiser@dartmouth.edu}

\author{Sara Imari Walker}
\email{sara.i.walker@dartmouth.edu}

\affiliation{Department of Physics and Astronomy, Dartmouth College
Hanover, NH 03755, USA}

\begin{abstract}
The activation-polymerization-epimerization-depolymerization (APED) model of Plasson {\em et al.} has recently been proposed as a mechanism for the evolution of homochirality on prebiotic Earth. The dynamics of the APED model in two-dimensional spatially-extended systems is investigated for various realistic reaction parameters. It is found that the APED system allows for the formation of isolated homochiral proto--domains surrounded by a racemate. A diffusive slowdown of the APED network  induced, for example, through tidal motion or evaporating pools and lagoons leads to the stabilization of homochiral bounded structures as expected in the first self-assembled protocells. 
\end{abstract}


\maketitle

\begin{flushleft}
Abbreviations: APED, activation-polymerization-epirmerization-depolymerization; e.e., enantiomeric excess; NCA, N-caboxyanhydride
\end{flushleft}

\section{Introduction}

While much progress has been made in understanding the many aspects related to the origin of terrestrial life, a coherent framework of events leading up to first life is still beyond reach. Current theories of the origin of life typically address one of its aspects. For example, investigations may focus on prebiotic chemistry as in the Miller experiment \cite{Miller} and related follow-up studies \cite{Rode} or on the formation of the first cellular membranes as in the work of Deamer and collaborators \cite{Deamer, MBD}. Here we attempt to combine a recent scenario for the onset of homochirality in peptide systems with models describing the formation of localized cell-like proto-domains in an effort to synthesize some of the ideas presented in the literature into a cohesive framework.

A critical step in the emergence of life was the advent of cellular membranes. Membranous boundaries are a key component of all modern cells and are required for processes such as speciation, energy capture, and metabolic catalytic reactions \cite{Monnard}. ``Minimum protocells'' - membrane vesicles capable of self-replication - have been proposed by Morowitz {\it et al.} as a possible first step toward functioning cells fitted for self-replication and Darwinian evolution \cite{MBD}. The appeal of this membrane-first, lipid-world scenario is the ability of such boundary structures to thermodynamically separate their contents from the external environment: the interior of the membrane provides a closed microenvironment for chemical reactions while the exterior environment provides free energy and nutrients \cite{MBD}. This view is supported by numerous studies focusing on the formation of lipid membrane-vesicles from prebiotically relevant ingredients: vesicle formation has been demonstrated for a variety of amphiphilic compounds \cite{Deamer, Deamer2, Monnard, LLWR, BMWL, BBML} including formation of boundary structures from organic molecules found in the Murchison meteorite \cite{Deamer3}.

In contrast to the perspective provided by proponents of the lipid-world scenario \cite{Segre}, self-organization of short peptides has recently been proposed as a possible alternative to lipid self-organization for formation of the first protocellular membrane structures \cite{Fishkis, Santoso}. Although phospholipids are the primary components of modern cellular membranes, it is unlikely that such compounds were prevalent enough on the early Earth to be major components of the very first membranes \cite{Deamer, Lazcano}. As pointed out by Fishkis \cite{Fishkis}, the first cellular membranes may have consisted of polypeptides, possibly in conjunction with other prebiotic amphiphilic molecules.  
This idea is supported by findings indicating that short peptides can aggregate into fibrils, microtubes, and vesicles \cite{Reches, Reches2, Santoso, Vauthey}.

The notion of a ``peptide-first'' origin of life where polypeptides self-organize into protocellular structures is not new. Earlier work by Fox and coworkers \cite{Fox, Fox2} addressed the formation of proteinoid microspheres from self-organizing amino acids. It was found that heated amino acids can self-order in copolymerization reactions and self-organize when in contact with water to form protocells (also referred to as minimal or pre-contemporary cells \cite{Fox}). These protocells can feature many of the properties common to life: metabolism, growth, reproduction and response to stimuli in the environment \cite{Fox3}. They are also capable of evolving to more modern cells - including the initiation of a nucleic acid coding system \cite{Fox3}. However, the polypeptides in these studies are composed of mixed $L$-- and $D$--amino acids with no mention of how a microsphere might evolve toward a homochiral contemporary cell with exclusively $L$--amino acids \cite{Fox}.

Apart from studies of the origins of the first cellular boundaries, the origin of life's chiral signature, {\em i.e.} dextrorotatory sugars and levorotatory amino acids, has been a longstanding question in studies of abiogenesis \cite{Bonner}. The first breakthrough in understanding biomolecular chiral symmetry-breaking occurred with the influential work of Frank \cite{Frank53}, where autocatalysis and some form of mutual antagonism were identified as sufficient ingredients for obtaining molecular homochirality.  Since then, many ``Frank models'' have been proposed, each providing its own description of chiral symmetry breaking.  These models range from the early work on bifurcation models by Kondepudi and Nelson \cite{KN83} and Avetisov and Goldanski \cite{AG93} to more recent investigations of crystallization \cite{SH05, Viedma}, and chiral selection in polymerization \cite{Sandars03, SH052}.  One of the better known models is that of Sandars \cite{Sandars03, BM, WC, GW}, where the mutual antagonism is provided by enantiomeric cross-inhibition in template-directed polycondensation of polynucleotides \cite{Joyce}. Although this model provides a theoretically elegant description of how homochirality might emerge in a RNA world, the autocatalytic reactions necessary for chiral symmetry breaking to be observed in such a system are presently only demonstrated for a few non-biological molecules \cite{Blackmond04, Soai} and would be trying to achieve with even very simple organic molecules \cite{Joyce2}.

A more recent proposal, describing the emergence of homochirality in an early peptide world has been provided by Plasson, Bersini, and Commeyras \cite{Plasson04}. A prebiotically relevant reaction network consisting of activation, polymerization, epimerization, and depolymerization (APED) of chiral peptides was shown to allow self-conversion of all chiral subunits to a single handedness. As discussed by Plasson {\em et al.}, and later by Brandenburg {\em et al.} \cite{BLL}, the APED model is distinct from the class of Frank models in that symmetry breaking in APED {\em mimics} autocatalysis  
without any of the molecules actually possessing catalytic behavior. The APED system therefore represents what may be the first in a class of models which are not dependent on autocatalysis - a beneficial feature given that direct autocatalysis is not believed to be possible with short nucleotides or short peptides \cite{BLL}. 

Another attractive feature of the APED model is that it describes a mass--conserving, closed, recycled system of reversible chemical reactions allowing complete conversion of all chiral subunits to a single chirality: in other words, with continual input of energy, the APED system provides a description of {\it emergent} chirality, where one may begin with a closed near-racemic system of left and right-handed deactivated monomers and end with a nearly homochiral system of monomers and dimers. It is this feature that makes the model appealing -- it provides a bottom--up approach to the onset of homochirality. 

A bottom-up approach to the origins of life, based on increased complexification from primordial prebiotic precursors \cite{Frank53,Miller,Lazcano, AG93, Sandars03, BM,GW,Rode} is usually contrasted with top-down methods aimed at reducing cells to the minimal ingredients necessary to still be deemed living \cite{Rasmussen, Szostak}. Although there is no doubt that both approaches are fundamentally sound and worth pursuing, very little work has been done in trying to bridge the gap between the two, that is, in trying to go from chemistry to biology. Our goal in the present manuscript is to take the first steps in this direction, by providing a workable closed chemical model which evolves dynamically through its own self-interactions towards precursors of cell-like homochiral structures surrounded by a racemic environment. By ``cell-like'' we mean spatially-bounded structures with contents differentiated from the exterior and with a hint of proto-metabolic activity. We do not attempt here to model full cell-like behavior, including transmembrane ionic transport activity and division with transmission of genetic information. Our goal is to provide the possible first steps toward more complex and realistic cell-like behavior including homochiral biochemistry.

For this end, we will introduce spatial dynamics to the APED model and obtain the reaction rates allowing for the emergence of localized homochiral domains.  We will see that both tube-like and spherical domains are possible, depending on the details of the dynamics. We begin our discussion with a description of some of the key features of the APED model in the section {\em The Model}. We then introduce spatial dynamics and discuss our findings in the {\em Results} section. In the {\em Discussion} we make contact with proto--cellular abiogenesis in light of the work by Fox and colleagues \cite{Fox, Fox2, Fox3} and by Fishkis \cite{Fishkis}, highlighting peptides as a potentially crucial ingredient in the first protocells. We also briefly discuss the consequences of our work to future searches of extraterrestrial stereochemistry.

\section{The Model}

The polymerization reactions in the original work of Plasson {\em et al.}  were limited to dimerizations \cite{Plasson04}: a relevant truncation given that dipeptides have been identified in the Murchison meteorite \cite{Meierhenrich}. We follow this simplification here, since solving the coupled network of reaction equations becomes quite complex  when spatial dynamics are included. The set of APED reactions limited to dimerizations is: 
\begin{eqnarray}\label{rxnNetwork}
L  \stackrel{a}{\rightarrow} L^*  &~ L^*  \stackrel{b}{\rightarrow} L \nonumber \\
L^* + L   \stackrel{p}{\rightarrow} LL & ~ L^* + D   \stackrel{\alpha p}{\rightarrow} LD \nonumber \\
LL   \stackrel{h}{\rightarrow} L + L &~ DL   \stackrel{\beta h}{\rightarrow} D + L  \nonumber \\
DL   \stackrel{e}{\rightarrow} LL  &~ LL   \stackrel{\gamma e}{\rightarrow} DL 
\end{eqnarray}
supplemented by the complementary reactions with $L \leftrightarrow D$. By convention \cite{Plasson04}, the reacting amino acid is on the left side of a polymer chain. Here $a$ and $b$ are the activation and deactivation rates of monomers, respectively. The rates $p$ and $h$ describe the respective rates of polymerization and depolymerization of homochiral dimers, while $e$ describes the rate of epimerization of heterochiral dimers. The rates $\alpha p$, $\beta h$, and $\gamma e$ then correspond to the complementary stereospecific reactions rates for polymerization and depolymerization of heterochiral dimers and epimerization of homochiral dimers, respectively, where $\alpha$, $\beta$, and $\gamma$ quantify the degree of stereoselectivity.
An important feature of the model is that the system is closed with respect to mass flux and maintains a constant total concentration of amino acids and peptides, 
\begin{eqnarray} \label{eqn:c}
c \equiv  [L] + [D] + [L^*] + [D^*] + 2([LL] + [DD] + [LD] +[DL])~.
\end{eqnarray}

Successful symmetry breaking in APED systems requires stereospecific reactions, {\em i.e.}, $\alpha$, $\beta$, $\gamma \neq 1$. Studying the case of complete stereoselectivity  for depolymerization and epimerization reactions ($\beta = \gamma = 0$), Plasson {\em et al.} found that the system allows four types of fixed points, labeled asymmetric, symmetric, dead, and unstable. The behavior of the system is thus governed by the nonzero stereoselectivity parameter, $\alpha$, and the conserved total mass through the parameter $c$. Together, these two parameters control the type of fixed point. In {\it all} cases, if $c$ is below a critical value, no dimers ever form. This is the (racemic) dead solution. The existence of these low concentration dead solutions becomes important when we study spatially extended systems. A stereoselectivity favoring formation of homochiral dimers ($\alpha <1$) was found to be essential in producing stable homochiral solutions \cite{BLL, Plasson04} even in the relevant case of increased stereoselective pressure due to the presence of nonzero $\beta$ and $\gamma$. Therefore, in this work we will restrict our investigations to the case $\alpha < 1$. 

As pointed out by Plasson {\em et al.}, the model can be applied to realistic systems of amino acid derivatives \cite{Plasson04}. Although amino acids have been shown to be synthesized under a variety of prebiotically relevant conditions \cite{Miller, Rode}, this fact alone is not enough to guarantee the eventual emergence of peptide chemistry or life -- free amino acids are notoriously poor reactants. Therefore, the limiting factor in primitive peptide synthesis is most likely {\em not} the formation of the amino acids but, instead, their eventual activation. 

The most commonly cited potential prebiotic activations of stable amino acid derivatives lead to the formation of $N-$carboxyanhydrides of $\alpha$-amino acids \cite{Com2002, Huber98, Leman}. NCAs have several features that make them highly amenable to building a desirable prebiotic chemistry ({\it i.e.} a {\em homochiral} chemistry) based on the APED set of reactions \cite{Plasson04}:
\begin{enumerate}
\item NCAs are very reactive amino acid derivatives, which easily permit the formation of peptides in aqueous solutions.
\item The polymerization of NCAs is stereoselective, favoring the formation of homochiral peptides.
\item The chemistry of NCAs promotes the recycling of products.
\end{enumerate}

There exists a wide range of models based on the chemistry of NCAs with different energy sources for activation. Potential compounds that may have acted as sources of energy to activate amino acids on the primitive Earth include nitrogen oxides, as in the primary pump model of Commeyras and coworkers  \cite{Boiteau, Com2002}; CO, familiar from the work of W\"achtersh\"auser and collaborators \cite{Huber98, Huber03, Wach}; or COS, as in the model of  Leman {\em et al.} \cite{Leman}. Such sources of chemical energy would have existed on the early Earth, possibly in large quantities due to increased geological activity during the Hadean era such as outgassing from volcanos. For a NCA-driven APED chemistry, these activating compounds provide the necessary free energy required for the system to have operated in a primordial environment \cite{Plasson04, Plasson08}.  

\section{Results}

The introduction of spatial dependence permits the net enantiomeric excess ($ee$) to evolve in space as well as time, thereby unveiling new and interesting dynamics in the APED system. We establish a spatiotemporal reaction-diffusion network from the rate equations derived from eqs. \ref{rxnNetwork} by following the usual procedure in the phenomenological treatment of phase transitions outlined below.  Systems with either spatially homogeneous and inhomogeneous concentrations are investigated separately. 

\subsection{Introducing Spatial Dependence} \label{SD}
Spatial dependence was introduced following the usual procedure in the phenomenological treatment of phase transitions \cite{Gunton, Langerrev} by rewriting total derivatives as $d/dt \rightarrow \partial / \partial t - k_{1,2} \nabla ^2$ in the rate equations derived from eqs. \ref{rxnNetwork}.  This established a reaction-diffusion network with diffusion constants $k_1$ and $k_2$ describing the molecular diffusion of monomers and dimers respectively. It is convenient to introduce the dimensionless diffusion constants $\kappa_{1,2} =  k_{1,2}/ \kappa$, where $\kappa$ is a dimensionful diffusion rate which we scale out of the network equations (typically we set $\kappa_1 = 2\kappa_2$). Some values of physically relevant diffusion rates are $\kappa = 10^{-9}$m$^2$s$^{-1}$ for molecular diffusion in water and $\kappa = 10^{-5}$m$^2$s$^{-1}$ for air \cite{Cott}. The corresponding rate equations were then made dimensionless by scaling all concentrations $[U]$ (with $U$ being any monomer or dimer in the network) such that the dimensionless concentrations are written as $u = \frac{p}{a} [U]$. This allowed us to scale the reaction rate $p$ out of the network equations and to write all other network reaction rates in terms of dimensionless ratios with $a$ (see below). Here we have chosen $a$ as our tunable parameter as it is dependent on activation by consumption of fuel compounds from the primordial environment and is therefore more flexible than other model parameters \cite{Plasson08}. We also introduce the dimensionless time and space variables $t_0 \equiv a t$ and $x_0 \equiv \sqrt{a/\kappa} ~~ x $. One may then recover dimensionful values for a particular choice of the rate $a$ and the diffusion coefficient $\kappa$. The dimensionless rate equations corresponding to eqs. \ref{rxnNetwork} (with $b=0$) were found to be, with $l \equiv  \frac{p}{a} [L]$, $d\equiv  \frac{p}{a}[D]$, etc.:
\begin{eqnarray}\label{eqns}
 \frac{\partial l}{\partial t_0} - \kappa_1 \nabla_0^2 l &=& -l-(l^*+\alpha d^*) l  + 2 \eta ll  + \beta \eta (ld + dl),
\nonumber \\
 \frac{\partial l^*}{\partial t_0} - \kappa_1 \nabla_0^2 l^*&=& l  - (l+\alpha d)l^*, \nonumber \\
 \frac{\partial ll}{\partial t_0} - \kappa_2 \nabla_0^2 ll &=& l^*  \cdot l - \eta ll + \varepsilon(dl - \gamma ll), \nonumber \\
 \frac{\partial ld }{\partial t_0} -  \kappa_2  \nabla_0^2 ld &=& \alpha l^*\cdot d - \beta \eta ld - \varepsilon(ld   - \gamma dd),
\end{eqnarray}
and the corresponding four equations obtained with $l \leftrightarrow d$. The dimensionless reaction rates $\eta$ and $\varepsilon$ are defined as $ \eta \equiv h/a$ and $ \varepsilon \equiv e/a$ and ${\scriptstyle \nabla_0^2}$ is the dimensionless Laplacian (in ${\scriptstyle 2d}$, ${\scriptstyle \nabla_0^2 = \partial^2 / \partial x^2 + \partial^2 / \partial y^2}$).  For the work presented here we have chosen values of $\kappa_1$ and $\kappa_2$ such that $\kappa_1 = 2 \kappa_2$. 

We also note that the global $ee$, defined by Plasson {\em et al.} as a measure of the net asymmetry, is given by 
\begin{eqnarray} \label{eqn:ee}
ee = ( [L] +   [L^*] + 2[LL] - [D] - [D^*] - 2[DD] )/c~,
\end{eqnarray}
where $c$ is given in eq. \ref{eqn:c}. The net enantiomeric excess we cite in this work is then found by spatial averaging, {\em i.e.}, 
\begin{eqnarray} \label{eqn:eea}
\langle ee \rangle = (1/A) \int (ee) d^2 x~, 
\end{eqnarray}
where $A$ is the area.

\subsection{Numerical Simulations with Stochastic Initial Conditions} \label{NS}

The numerical implementation utilized a finite-difference leapfrog algorithm with spatial step $\delta x = 0.2$ and temporal step $\delta t = 0.005$. Periodic boundary conditions were adopted on a $512^2$  square lattice in two dimensions.  Near-racemic initial conditions were prepared in a two step process. First, a racemic lattice configuration was established using the analytic form of the symmetric steady-state solution to the network equations in eqs. \ref{eqns} (with $b=0$), found as a function of the total concentration $c$, given in eq. \ref{eqn:c}. The dependence of the analytic solution on the total mass allowed us to use the distribution of $c({\bf x})$ over the lattice to determine whether we would study a homogeneous or inhomogeneous system. For homogeneous systems, we set a uniform concentration $c({\bf x}) = \langle c({\bf x}) \rangle > c_{\rm crit} = 1.5$ such that symmetry breaking would be observed everywhere in the system (here $c_{\rm crit}$ is given for the case $\alpha = 0.35$). For inhomogeneous systems, initial high concentration regions ({\em i.e.} regions where $c > c_{\rm crit}$), or bubbles, were established with Gaussian profiles such that $c({\bf x}, 0) =  {\cal C}\exp[-(x^2 + y^2)/R^2]$ where ${\cal C}$ is the total concentration of amino acids and peptides at the bubble's core (${\cal C} > c_{\rm crit}$), and $R$ is the radius of the bubble. Outside of the bubbles, the concentration was set to zero. The final step was to create stochastic initial conditions via a generalized spatiotemporal Langevin equation. We added a Gaussian white noise term, $\xi(x, y, t)$, with zero mean and two-point correlation function $\langle \xi({\bf x'}, t') \xi({\bf x}, t) \rangle = \theta^2 \delta (t' -t)\delta ({\bf x'} -{\bf x})$, to the rate equations. The equations of motion, $u(x, y, t)$ from eqs. \ref{eqns} with the added noise terms were then solved with simple quadratic potentials, $V(u) = (1/2) u^2$ and low noise ($\theta^2 = 0.002$) until the volume-averaged enantiomeric excess achieved a steady-state Gaussian distribution centered on the racemic state. By steady-state we mean $d \langle ee(t) \rangle /dt \approx 0$, where $\langle ee \rangle$ is given by eq. \ref{eqn:eea}. Once the near-racemic initial condition was established, the noise ($\xi(x, y, t)$) was turned off and the system was then set to evolve deterministically. 

\subsection{Homogeneous Systems and Chiral Domains} 

\begin{figure}
\centerline{\includegraphics[width=3.5in]{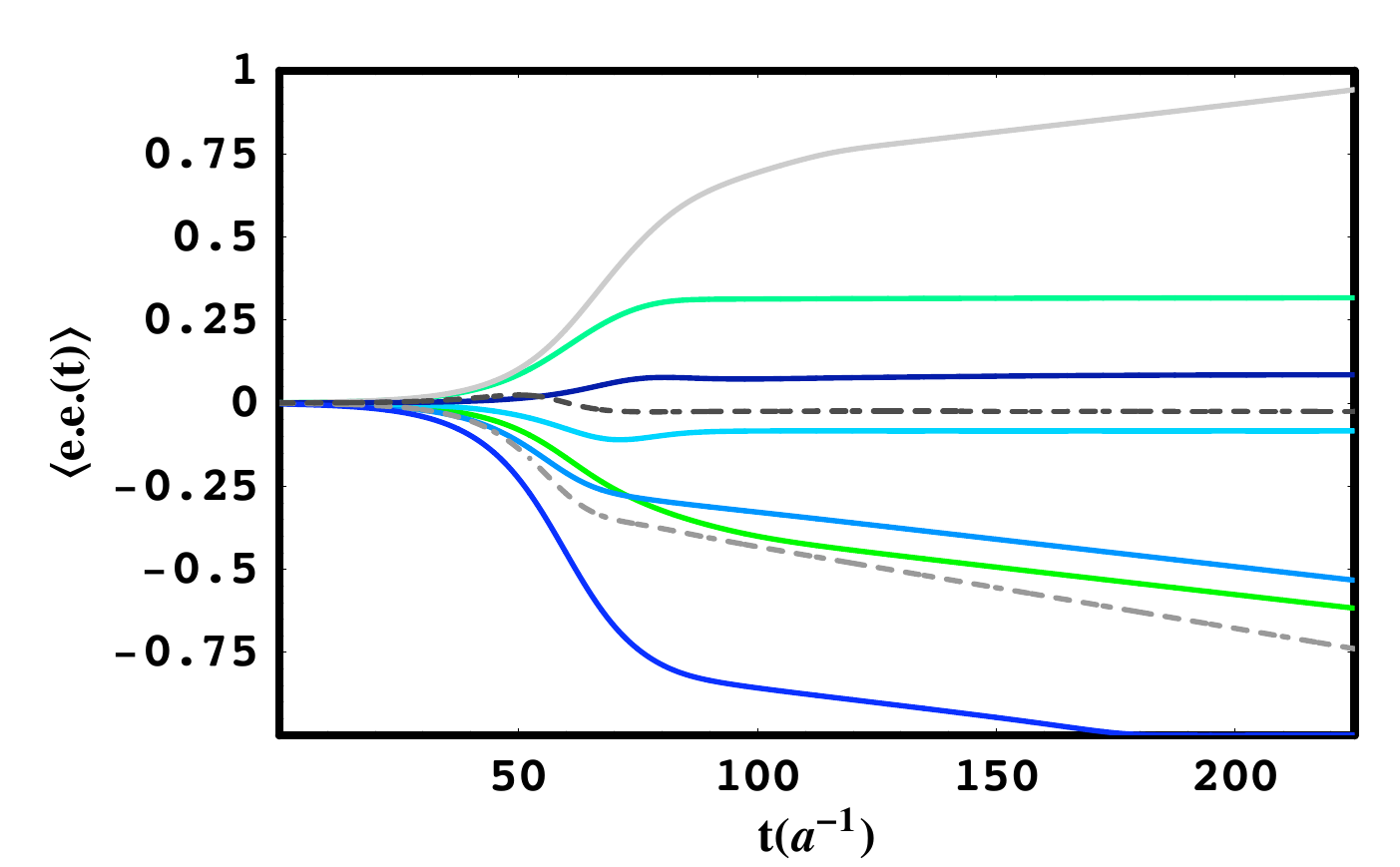}}
\caption{Time evolution of spatially-averaged  ee in homogeneous APED systems. Each line represents a system starting from a random near-racemic distribution ($ee(t=0) < 0.001$) of monomers and dimers. The two dashed lines correspond to the systems shown in Figure \ref{fig:Fig2}  (the darker dashed line corresponds to the top panel and the lighter line to the bottom panel).} \label{fig:Fig1}
\end{figure}

\begin{figure}
\centerline{\includegraphics[width=5in]{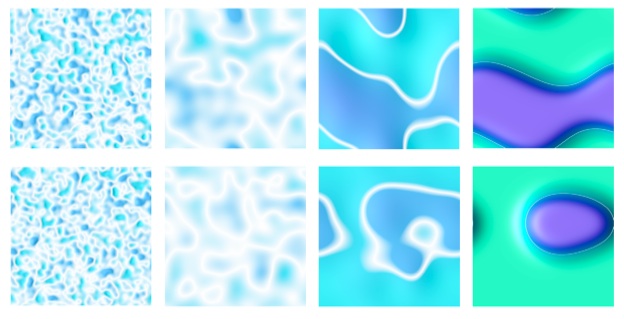}}
\caption{Evolution of chiral domains. Time increases from left to right. {\bf Top}. Formation of two percolating domains of opposing chirality. The phases of $L$ (dark blue) and $D$ (aqua) achieve near complete homochirality divided by a thin racemic (white) domain wall. {\bf Bottom}. Only one domain (aqua) percolates and the system evolves toward a single chirality ($ee = -1$, corresponding to the homochiral $D$-phase).} \label{fig:Fig2}
\end{figure}

We first consider the formation of chiral domains in systems with homogeneous concentrations ({\em i.e.} $c(\textbf{x},t)=$ constant).With a near-racemic initial distribution, a typical spatial volume ${\cal V} \gg \lambda^2$, where $\lambda$ is the correlation length, will coarsen into domains of left and right-handed chiralities \cite{BM,GT,G}. This occurs after a long induction period where activated monomers of both chiral species accumulate in the system. During the induction period, the $ee$ (given in eq. \ref{eqn:ee}) remains small; however, once a sufficient concentration of activated monomers has been achieved we begin to observe regions of growing enantiomeric excess. Results demonstrating the evolution of $\langle ee \rangle$ are shown in Figure \ref{fig:Fig1} for a number of runs (the dashed lines in the figure correspond to the two sample systems of Figure \ref{fig:Fig2}).  We present data for both the unrealistic case of complete stereoselectivity of depolymerization and epimerization reactions ($\beta = \gamma = 0$) and for more realistic nonzero parameters ($\beta$, $\gamma \neq 0$), although for the homogeneous systems of this section we restrict our analysis to $\beta=\gamma=0$. We note that the results are qualitatively similar even if with realistic parameters the final $ee$ never reaches a complete homochiral state (see \cite{Plasson04}). As discussed in the introduction, other polymerization models describing the onset of homochirality yield similar results for the evolution of the spatially averaged $\langle ee \rangle$ \cite{GW} without the activation period which is typical of the APED model. 

We have observed two types of domain evolution, as shown in Figure \ref{fig:Fig2}. In the top row, domains of both chiralities percolate ({\it i.e.} permeate the entire volume) through the volume and the system reaches a steady-state consisting of two domains with opposing chirality separated by a thin domain wall. In the bottom row, only one chiral domain percolates and the system eventually evolves toward homochirality. The parameters chosen here ($\beta=\gamma =0$ with $\alpha = 0.3$, $\varepsilon = \eta = a$, and $b = 0$) allow for near complete conversion to homochirality: within a single domain, the net $ee\geq 99 \%$. 

\begin{figure}
\centerline{\includegraphics[width=5in]{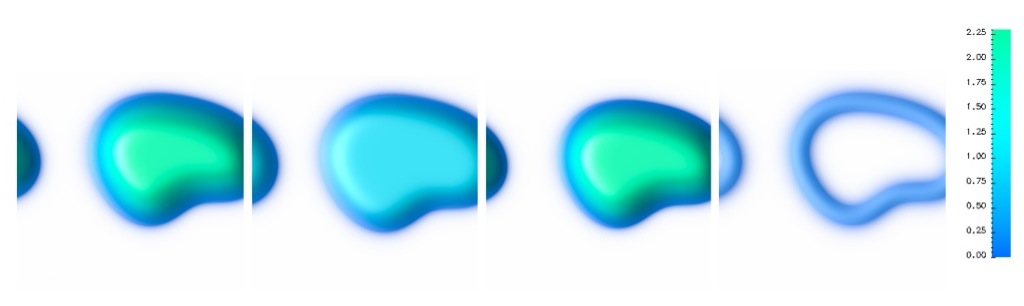}}
\caption{Snapshots of molecular concentrations at one instant in time. From left to right, the snapshots show concentrations $L$, $L^*$, $LL$, and $LD$. White denotes regions where the concentration of a molecular species is $< 0.15$ (these regions are where an excess in $D$-enantiomers resides). Note that the near homochiral domain of the $L$-phase is surrounded by a racemic ring of $LD$ and $DL$ (not shown) forming a boundary around the domain.} \label{fig:Fig3}
\end{figure}

Quite interestingly, the domain wall or ring-like structure surrounding the  homochiral domains consists of a racemic mixture of monomers and dimers including large concentrations of the heterodimers $DL$ and $LD$, as shown in Figure \ref{fig:Fig3}. This is very suggestive of a physical barrier isolating the inside of the homochiral domain and is true both when only one chirality percolates (bottom-right snapshot), and when both chiralities percolate (top-right snapshot), where one observes the formation of tubular domains. These situations are  suggestive of micelle vesicles and bilayer membranes, observed in emerging self-assembled structures of amphiphiles \cite{Deamer}. It is thus of great interest to examine the possibility of assembling similar structures in an inhomogeneous context, such that homochiral protodomains may become surrounded by a racemate.

\subsection{Inhomogeneous  Systems: The Role of Non--Uniform Concentration}

\begin{figure}
\centerline{\includegraphics[width=5in]{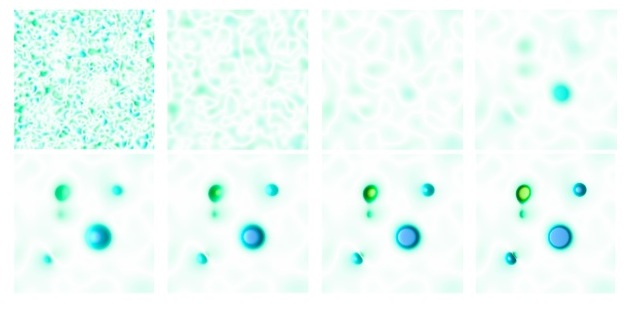}}
\caption{Evolution of chiral protocell-like regions. Time increases from left to right and top to bottom. The diffusion was sharply decreased ($\kappa_2 = 0.1$ to $\kappa_2 = 10^{-6}$) between the top and bottom rows.  {\bf Top Row}: $t_0 = 3, 15, 45, 75$. Racemic regions (white) grow as activated monomers build in the system, followed by the appearance of proto-domains of chiral excess in regions where $c > c_{crit}$. {\bf Bottom Row}:  $t_0 = 135, 165, 195, 240$. The proto-domains continue to coarsen within the racemic (white) background as the magnitudes of the $L$ (dark blue) and $D$ (yellow-green) phases increase (note that since $\beta, \gamma \neq 0$, complete homochirality is not possible Ð the maximum magnitude of the $ee$ in this system is $|\langle ee \rangle | = 0.6$, where $\langle ee \rangle$ is defined in eq. \ref{eqn:eea}. } \label{fig:Fig4}
\end{figure}

The examples discussed thus far have maintained a uniform concentration over the entire lattice. We now consider the more realistic dynamics of systems with spatially-dependent concentrations ($c(\textbf{x},t) \neq $ constant). In this case, starting from stochastic near--racemic initial conditions, regions where chiral symmetry breaking occurs and regions where a racemate is maintained should emerge dynamically. Holding all reaction parameters fixed, the previous results indicate that we should observe this separation in systems where some regions have concentrations above and others below a critical, parameter-dependent value (see section {\em The Model}). Accordingly, our simulations show that the spatially-extended APED system naturally develops chiral protocell-like structures surrounded by a racemate, as shown in Figure \ref{fig:Fig4}.

Left to evolve diffusively, the high-concentration protocellular domains containing an enantiomeric excess (Figure \ref{fig:Fig4} lower panel) would eventually shrink away. However, we must recall that in prebiotic Earth reactor pools were submitted to environmental disturbances ranging from mild (e.g. tides, evaporating lagoons) to severe (e.g. volcanic eruptions, meteoritic impacts). Both kinds of disturbances affected the evolution of chirality in early Earth \cite{GTW}. Recent studies of life's origins have focused on molecular interactions in the context of changing tides \cite{Boiteau, Bywater, Com2002, Lathe1, Lathe2} or evaporating pools and lagoons \cite{RM} as ways to enhance the concentrations of reactants. From a modeling perspective, their effect can be mimicked by a time-dependent modulation in the diffusion rates. 

To simulate such processes, we started the system's evolution with a fast diffusion coefficient ($\kappa_2 = 0.1$, corresponding to molecular diffusion slightly slower than that in water with dimensionful values) followed by evolution with a very slow diffusion rate ($\kappa_2 = 10^{-6}$) such as might occur for molecules diffusing on dry clay, a rocky surface, or on a beach. The initial conditions were set such that the lattice had, on average, a low concentration of amino acids and peptides: $\langle c({\bf x}) \rangle < c_{\rm crit}$, where $c_{\rm crit} = 1.5$ for the systems studied. Bubbles with higher concentrations, $ c({\bf x})  > c_{\rm crit}$, were interspersed on the lattice (see section \ref{NS}). The results of a sample run are shown in Figure \ref{fig:Fig4}. The stereoselectivities have been set such that $\alpha = 0.35$, $\beta = 0.2$, and $\gamma = 0.3$ as in the ``real system'' parameters from Plasson {\em et al.} \cite{Plasson04}, with the non-stereoselective reaction rates set to the dimensionless values $\varepsilon = \eta  = 10$ and $b=0$ ($a$ and $p$ have been scaled out of the network equations - see section \ref{SD}). Since the epimerization rate on the $N$-terminal is of the same order as the hydrolysis rate for NCAs \cite{Plasson04}, we have set $\varepsilon = \eta$ ({\em i.e.} $e = h$).

The top left snapshot in Figure \ref{fig:Fig4} shows a near-racemic distribution with regions of higher reactant concentration barely visible. In this simulation, the tide would ebb or the lagoon evaporate between the two leftmost snapshots in the bottom row.  Following the drying of the tidal pool or lagoon, regions with total reactant concentration initially above critical are seen to coarsen into domains with high $\langle ee \rangle$ surrounded by a near-racemic environment. 

\begin{figure}
\centerline{\includegraphics[width=5in]{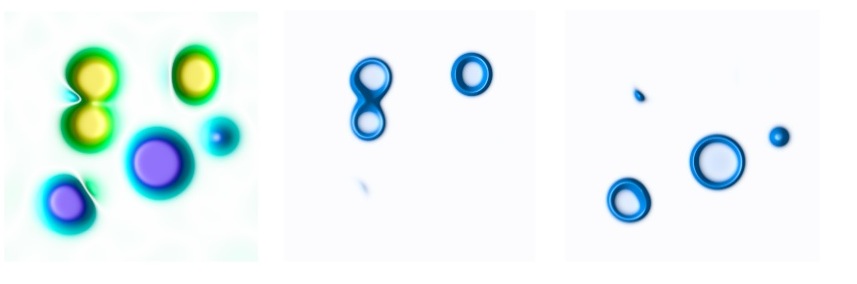}}
\caption{Snapshots of the enantiomeric excess and molecular concentrations of heterodimers at one instant in time. From left to right the snapshots show the $ee$, and the concentrations of $LD$ and $DL$. The color scale for the first snapshot is the same as that for Figure \ref{fig:Fig4} (note that the near-homochiral domains achieve larger $ee$ as compared to those of Figure \ref{fig:Fig4} due to the smallness of $\beta$ and $\gamma$). For the second two snapshots, white denotes regions were the concentration is $< 0.001$.} \label{fig:Fig5}
\end{figure}

As with the homogeneous case, we also investigated if the near-homochiral domains were surrounded by a ring-like structure rich in heterodimers LD and DL. In Figure \ref{fig:Fig5}, we see that, indeed, such rings do appear. The net concentrations are sensitive to the parameters $\beta$ and $\gamma$ and tend to be suppressed for high values. The figure shows a simulation with the values $\beta=0.001$ and $\gamma=0.001$. Also, as in Figure \ref{fig:Fig4}, the diffusion was sharply reduced to simulate a tidal ebb: in this case, the tide would ebb prior to the snapshots shown in Figure \ref{fig:Fig5}. Our results show that for small nonzero values of $\beta$ and $\gamma$ (typically $< 0.05$) the near-homochiral domains are surrounded by racemic boundary structures. 

\subsubsection{Toward Smaller Protodomains}

For suitable choices of our tuneable parameter $a$ -- a measure of the energy input into the reactor pool, we can recover physically relevant, dimensionful values for the timescale and spatial extent. Here we focus on the inhomogeneous case as it offers a realistic example of a spatially-extended prebiotic system. Taking $a = 10^{-8}$ s$^{-1}$ as in the real system parameters of Plasson {\em et al.}, the hydrolysis rate assumes the value $h =  10 a = 10^{-7}$ s$^{-1}$ indicative of either high or low pH conditions \cite{Plasson04, Smith}. Using the dimensionless space and time variables, $x_0 = 102.4$ and $t_0 = 225$ from the simulation, the physical space and time scales are $x = 35$ m  and $t = 713$ y, respectively, not unreasonable values for a small pond that after evaporating for hundreds of years leaves chiral domains behind. Increasing $a$ will {\it decrease} the physical size and timescales of the simulation. If, instead, we take $a = 10^{-6}$ 
s$^{-1}$, we obtain $x = 3.2$ m and $t = 7.1$ y, indicating faster time scales for evaporation of smaller lagoons or pools. A changing tide on the order of hours in microscopic protocellular domains is harder to achieve with presently accepted values of the reaction parameters.  However, we have completed additional simulations on $128^2$ square lattices using parameters $dt=0.0001$ and $dx=0.05$ where proto-domains with diameters of centimeters and timescales of years were achieved for $a= 10^{-5}$s$^{-1}$. Please note that these dimensions are not due to fundamental restrictions on the chemistry involved but due to the limitations of our numerical simulations.

\section{Discussion}

Our results demonstrate that for suitable parameter choices, the APED model exhibits chiral symmetry breaking leading to homochiral (or near-homochiral) domains in spatially extended systems. The results are qualitatively similar to those for other bottom-up polymerization models describing the onset of homochirality where domains of opposite chirality form and compete for dominance \cite{BM,GW}, although here the model allows for the possibility of complete conversion of all chiral subunits for suitable parameter choices. A key point is that symmetry breaking in APED systems is sensitive to the total mass: processes, environmental or otherwise, which acted to accumulate prebiotic compounds would have favored the emergence of homochirality. 

As discussed in the introduction, self-organization of short peptides has previously been proposed as a possible mechanism for formation of the first protocellular membranes \cite{Fishkis, Fox, Fox2}. In the context of the results presented here, this implies that in regions where homochiral peptides are formed, one may observe spontaneous emergence of peptide vesicles. In fact, a compelling feature of polypeptide self--organization is that homochiral polypeptides have the propensity to form $\beta$-sheets while racemic mixtures will not exhibit such structural conformation \cite{Brack}. For molecules consisting of homochiral and heterochiral segments, structural formation entails aggregation of stable homochiral $\beta$--sheets surrounded by less stable and disordered heterochiral segments \cite{Brack}; furthermore, when subjected to mild hydrolysis, the more stable homochiral sheets are enriched by the residual fraction. These mechanisms may play a significant role in structural formation within the near--homochiral regions shown in Figure 3. It is therefore plausible that formation of $\beta$-sheets will be enhanced in homochiral regions, consistent with favoring vesicle formation in regions with high $ee$, although there is no experimental proof of this correlation. In this scenario, homochiral, high concentration regions are preferentially selected to evolve into primitive protocellular structures.

Self--organized homochiral protocellular structures could potentially share many features in common with modern cellular life. Encapsulation of macromolecules has been discussed in the context of lipid membranes via dehydration--rehydration processes \cite{Deamer}; it is possible that such processes may have operated in peptide membranes as well \cite{Fishkis}. Short peptides have been shown to exhibit catalytic activity under prebiotic conditions \cite{Shen}. In addition, a potential energy source for the protocell could be the chemical energy of activated amino acids \cite{Fishkis} that would still be present in the system at steady state.  In fact, the APED model has been described as a protometabolic pattern \cite{Plasson04}. At the very least, peptide protocells represent feasible prototypes displaying rudimentary features of modern cellular life \cite{Fox, Fishkis}. In light of the work presented here, this may now be extended to include homochiral protocells in the context of the APED model. We stress that our work says nothing of more complex cell behavior, such as nutrient exchange across cellular membranes and cell division. But it does provide a framework to study the emergence of isolated homochiral domains within a promising peptide model, possibly a first step toward the formation of rudimentary cell-like structures in early Earth.

We also note that in the absence of a specific chiral bias, the net enantiomeric excess in the APED model, and in many models of chiral symmetry breaking, is random and may be attributable to amplification of stochastic fluctuations about the initial near-racemic state \cite{Dunitz} due to environmental influences \cite{GTW} which may or may not induce turbulence \cite{BM}. Therefore, as shown in Figure \ref{fig:Fig4}, isolated regions can develop opposite chirality. This raises the possibility of having life forms of opposing chirality competing for resources on the early Earth \cite{BM}. Extending this scenario to other potentially life-bearing planetary platforms, and consistent with the difficulties of using parity violation in the weak interactions as a consistent chiral bias \cite{G}, our results suggests the existence of stereochemistry of opposing chirality, and possibly life, elsewhere in the universe \cite{Castelvecchi, GTW}. It is hopeful that proposed robotic missions such as the Titan Organics Explorer, if funded, will be equipped to analyze enantioenriched organic compounds \cite{Lunine} shedding light on this fundamental question.

\vspace{0.5in}
This work was supported in part by a National Science Foundation Grant PHY-0757124. We thank R. Plasson and Axel Brandenburg for many interesting suggestions and comments. SW thanks NORDITA in Stockholm for their kind hospitality during the initial stages of this work.

\end{document}